\documentclass[10pt,conference]{IEEEtran} 
\usepackage{amsmath}
\usepackage{graphicx}
\usepackage{color}
 \usepackage[table]{xcolor}
\usepackage{subfigure}
\usepackage{balance}

\title{Multi-level Memristive Memory with Resistive Networks }

\author{\IEEEauthorblockN{Aidana Irmanova, and Alex Pappachen James }
\IEEEauthorblockA{Department of Electrical and Electronics Engineering\\
Nazarbayev University,
Astana\\
Email: http://www.biomicrosystems.info/alex}}

\begin{document}

\maketitle

\begin{abstract}

Analog memory is of great importance in neurocomputing technologies field, but still remains difficult to implement. With emergence of memristors in VLSI technologies the idea of designing scalable analog data storage elements finds its second wind.  A memristor, known for its history dependent resistance levels, independently can provide blocks of binary or discrete state data storage. However, using single memristor to save the analog value is practically limited due to the device variability and implementation complexity. In this paper, we present a new design of discrete state memory cell consisting of sub-cells constructed from a memristor and its resistive network. A memristor in the sub-cells provides the storage element, while its resistive network is used for programming its resistance. Several sub-cells are then connected in parallel, resembling potential divider configuration. The output of the memory cell is the voltage resulting from distributing the input voltage among the sub-cells. Here, proposed design was programmed to obtain 10 and 27 different output levels depending on the configuration of the combined resistive networks within the sub-cell. Despite the simplicity of the circuit, this realization of multilevel memory provides increased number of output levels compared to previous designs of memory technologies based on memristors. Simulation results of proposed memory are analyzed providing explicit data on the issues of distinguishing discrete analog output levels and sensitivity of the cell to oscillations in write signal patterns.
\end{abstract}


\section{Introduction}

Majority of the memory technologies in the market today can store two logic states reflecting the widespread need and use of binary logic of computing \cite{digitm}. Therefore the data obtained from sensors that are analog in nature has to be digitized before it is processed and stored in the memory \cite{adc} consuming time, energy and die area \cite{essentialsmemory}. An alternative approach would be to process and store the signals in analog domain, which requires traditional memory technologies to be exploited beyond binary logic.

There are several reasons to make an effort towards developing non-binary memory devices. One of them is the evolution of many valued logic gates that push the speed limits of computing \cite{malinowski2007many}. The other motivation of designing analog memory is that it can be used in neuromorphic designs, as it is useful for synaptic-weight storage in artificial neural networks \cite{AMsurveyNagy}. One of the early designs that exploited analog memory in neural networks was Adaline (Adaptive Linear Neuron or later Adaptive Linear Element) which memory element consisted of memistors \cite{anderson2000talking}. A memistor (memory resistor) is a three-terminal device, made from an electroplating cell, that had thousands of possible analog storage levels \cite{memistor}.The main limitation of the technology was its non-scaling property, and due to this, it was eventually abandoned. 

Another three-terminal device that was widely used to implement analog memory cells is a floating gate transistor. Analog output of floating gate memory devices heavily relied on accurate control of charge injections and operated in high range of voltage levels \cite{HRNAMC}. Since the state and dynamics of the device are directly controlled by pulsatile inputs, this similarity with neuron's behavior was exploited and described in \cite{fujita1993floating}. In \cite{lee1991analog} floating gate devices were used for building analog data storage of synaptic weights.  

Multi-level Flash cells \cite{lee2006multi} and  PRAM \cite{lee2009multilevel} are another set of attempts to build analog memory. Multi-level flash is also a floating gate device that uses multiple levels per cell that can store more than one bit using the same number of transistors. The use of flash memory in neural network architecture was described in \cite{khan2013divided, strukov2015memory}. PRAM is a type of non-volatile random-access memory that exploits behavior of chalcogenide glass \cite{simpson2011interfacial}. PRAM's switching time and inherent scalability makes it appealing element for building analog data storage, but its temperature sensitivity presents a notable issue \cite{lee2010phase}, that should be taken into account during fabrication. PRAM for brain-like associative learning is described in \cite{eryilmaz2014brain}

There are also emerging memory technologies such as based on magnetic devices that are claimed to achieve significantly high integration with GP SIMD (General purpose single instruction, multiple data) and utilizes resistive crossbar technology \cite{morad2016resistive}. A different approach to realise the resistive multilevel memory would be using a memristor, a two terminal electrical component, the value of which is the linear relation between the charge and flux  \cite{chua1971memristor, rabbani2015multilevel, mostafa2016process}. Memistor has multilevel resistance property that can be controlled by the voltage or current applied across the device as a function of time and can be programmed to any resistance level between a maximum possible resistance and minimum resistance \cite{kannan2015modeling}.

To develop a memristor based multilevel memory, the earlier designs widely used reference resistance arrays \cite{kim2010memristor}, crossbar technology \cite{sahebkarkhorasani2015non, duan2014analog}, hybrid structure with Complementary Metal-Oxide Semiconductors \cite{rabbani2015multilevel}. 
In this paper, we propose a novel design of a memory cell that exploits the multilevel property of memristors. The memristors are programed to ternary states, using a simplistic memristor circuit in a potential divider configuration, where the potential differences across the individual memristor states are controlled to generate multiple states that can be combined to achieve discrete analog states. 


The Section II explains the design of the proposed memory focusing on the number of stored values, setting the resistive states of memristive sub-circuits and simulation results of the proposed memory cell. Further, the analyses on discrete analog output levels are provided.

\begin{figure}[thpb]
\centering
\includegraphics[width=4cm]{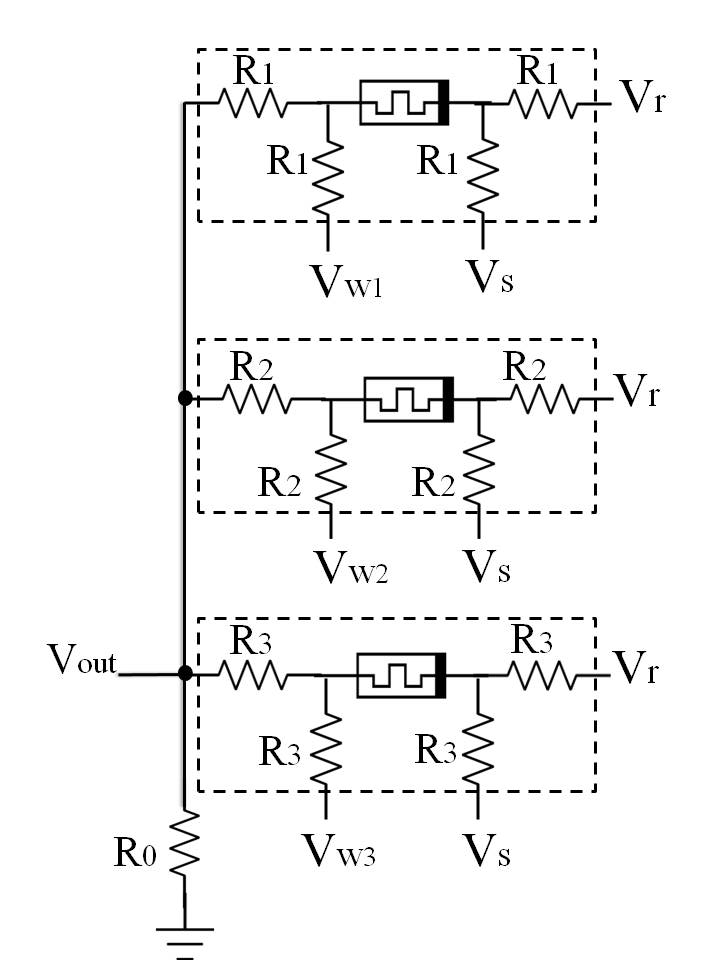}
\caption{Circuit design of the proposed discrete analog memory cell.}
\label{read}
\end{figure}

\section{Proposed Resistive Memory Network}

Memristors can store two logic states reflecting its ability to switch between two distinct resistance levels \cite{williams2008we}. The use of memristor to store multiple state requires the number of resistance levels to increase to multiple levels. While there have been attempts to quantize the resistance levels, it also increases the complexity of the devices. {In this work, we propose a circuit configuration that uses a set of  memristors to build a multi-level memory cell.}

\subsection{Number of stored values}
Figure 1 shows the proposed design of a memory cell that has 5 input (reset, read and 3 write ports) and 1 output ports constructed from 3 sub-cells. Each sub-cell of the memory consists of a memristor and equal valued 4 resistors, which are regarded as resistive network. Resistive network serves to separate each sub-cell from others. Another function of resistors is to separate $V_{r}$ read, $V_{s}$ reset, $V_{w1}$, $V_{w2}$, $V_{w3}$ write and $V_{out}$ output ports in the cell. 

In general, the  memory cell with $n$ sub-cells can store  $m^n$  discrete values, where $m$ is the number of different levels of voltage that is applied through $V_{w1}$, $V_{w2}$, $V_{w3}$. To program each memristor to the desired state the write signal $V_w$ is applied at the positive terminal of the memristors and the reset signal $V_s$, that precedes every write operation to erase previous states, is applied from the negative terminal of the device. Read signal $V_r$ is applied for reading the $m^n$  combination of memory states at the output $V_{out}$ of the circuit.

In this paper we present simulation results of the memory cell with $n=3$ sub-cells, each of them programmable to $m=3$(0,1,2) states which results in up to 27 level discrete analog memory.



\begin{figure}[!ht]

\centering
	\subfigure[]
    {
	\includegraphics
    [ height=3cm ]{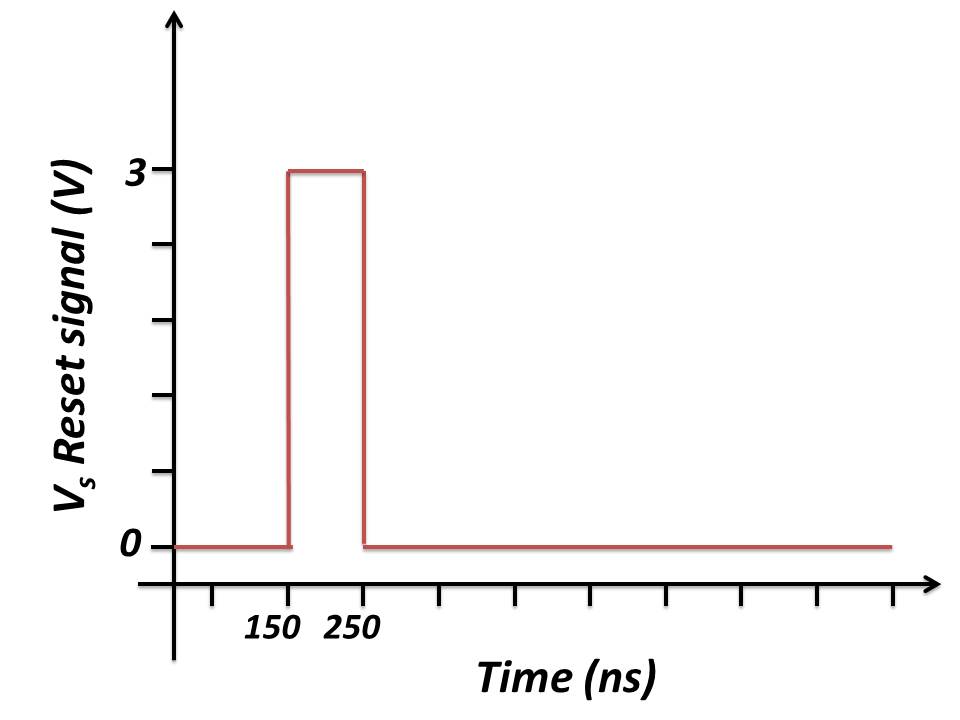}
	\label{l10}
    }
    \subfigure[]
    {
    \includegraphics[ height=3cm]{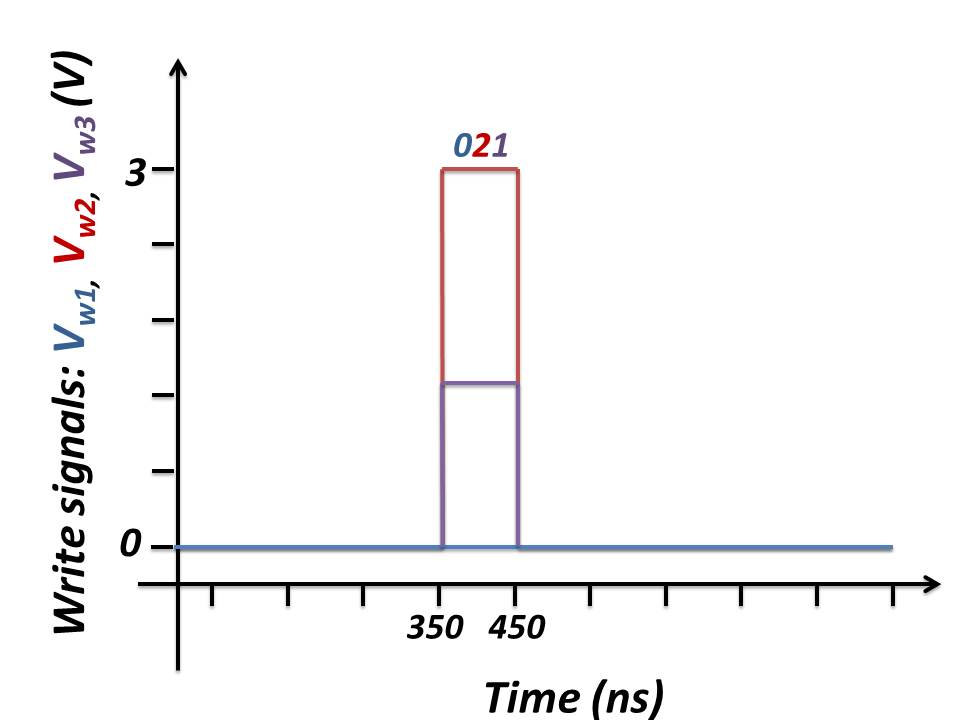}
  	 \label{l27}
     }
    \subfigure[]
    {
   \includegraphics[height=3cm]{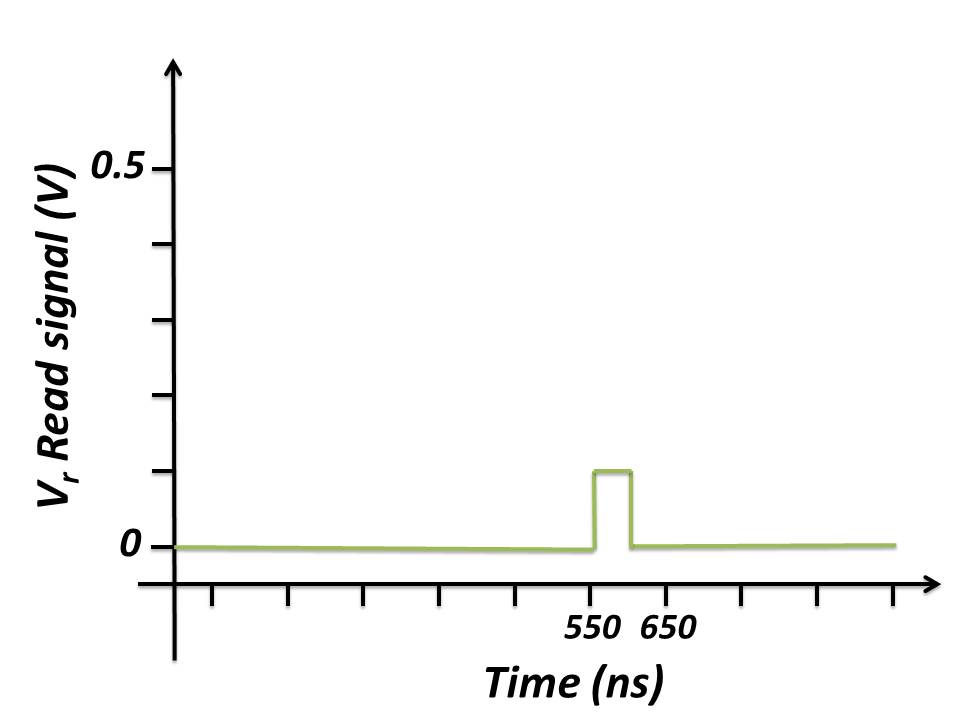}
   \label{l1024}
   }
     \caption{Timeline of the single cycle of (a) reset, (b) write and (c) read operations.}
     \end{figure}

\subsection{Setting the resistive states of a memristor }
The real-time programming of the memristor requires the application of a voltage pulse such that it is greater than its threshold switching voltage. The range of writing voltage can be selected according to the memristor model, so the highest positive values are for writing highest resistance levels and highest negative value is used for the reset operation. Throughout this paper, we have used Pickett memristor model \cite{kolka2015improved}, and kept the voltage in the range of -3V and +3V.

Implementation of the proposed design with three memristors in a single cell results in a multi-valued memory with ($3^3$=27) levels. Each sub-cell can store three different values indicating a ternary logic system with the possible $V_w$ values of $V_{max}$, $V_{max}/2$, and $V_{gnd}$ voltage levels (Figure 1).

The reset pulse signal with an amplitude of $V_{max}$ and time period of 100ns precede every write operation to restore the initial state of the sub-cell. In case of writing logic 0 write signal $V_w$ is set to 0V. The writing process itself is quite simple, for the desired value of logic 1 or logic 2  respective 1.5V, 3V voltage level is driven into the write line for the same 100 ns period. The write signal activates the cell and changes the resistive value of the memristor. Figure 2 illustrates the timeline example of writing the value of 021 to the cell overall, applying 0V, 3V and 1.5V to the $V_{w1}$, $V_{w2}$, $V_{w3}$ ports respectively, after resetting the sub-cells with single 3V $V_{s}$ pulse.

\begin{figure*}[!ht]
    \centering
      \subfigure[]{
    \includegraphics[height=4cm]{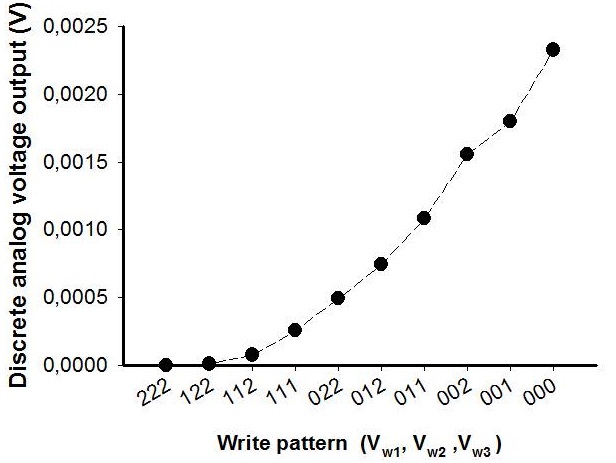}}
    \subfigure[]{
    \includegraphics[height=4cm]{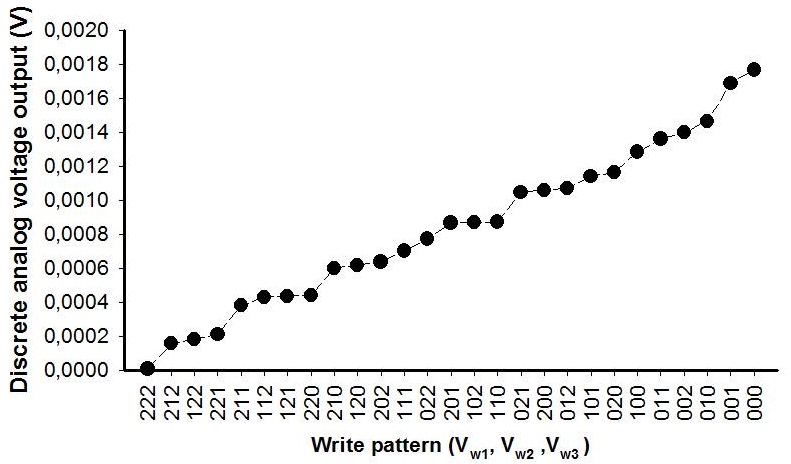}}
 
    \caption{The discrete analog levels for (a) 10 level memory cell, and (b) 27 level memory cell.}
\end{figure*}

\subsubsection{Read operation}

The read voltages $V_r$ of $0.1V$ that is much lower than $V_{max}$ is applied for shorter period - 50ns to enable the read operation without changing the resistive states of sub-cells. As it is shown in Fig.2 the control signal $V_s$ and the write enable signal $V_w$ are connected to the ground during the operation. The output voltage $V_{out}$ is indicative of the effective discrete analog level of the memory cell:



\begin{equation}
V_{out} = V_{r} \frac{\displaystyle\sum_{i=1}^{3}(\frac{1}
{R_{i1}k_{i}})}{(\displaystyle\sum_{i=1}^{3}\frac{1}{R_{i1}}(1-\frac{1}{k_{i}}) + \frac{1}{R_0})}
\end{equation}

where $n \in \{1,2,3\}$, $k_n = 2+\frac{R_{n1}}{R_{n2}}$, 
$R_{n1} =R_n+\frac{M_nR_n}{2R_n+M_n}$, and $R_{n2} = R_n+\frac{R_n^2}{2R_n+M_n}$.



\subsection{Simulations and results}

The simulation of the proposed circuit was done using HP memristor model \cite{kolka2015improved}. Configuring the resistive network as $R_1=R_2=R_3=20\Omega$ resulted in 10 output levels, but to achieve 27 output levels the values of resistors was set as $R_1=20\Omega$, $R_2=60\Omega$, and $R_3=180\Omega$. These semiconductor resistors account for the wire resistances and also help with differentiating voltage levels between the sub-cells. Figure 3(a) shows that output voltage levels resulting from simulation of the three memristor cells shown in Fig. 1 achieves 10 distinct discrete levels, when the resistors $R_1$, $R_2$ and $R_3$ set to same value of $20\Omega$. Changing the resistance values of the sub-cells i.e.  $\forall R_1\neq R_2 \neq R_3$ results in changing the voltage drop across the potential divider configuration within each sub-cells shown in Fig. 1. This change in output voltage of the sub-cells lead to differentiated voltage levels at the output of the proposed memory cell. Figure 3(b) shows an example of the increased number of discrete voltage levels to 27 when the resistor values are set to $R_1=20\Omega$, $R_2=60\Omega$, and $R_3=180\Omega$.

In Fig. 3(b) it can be noted that some of the values like 112 and 121, 121 and 220, 201 and 102, 102 and 110 have very close discrete voltages that can reduce the discrimination between the output voltage levels. On the other hand, when the wire resistance of the memristors are the same in each sub-cell (20 ohms) only 10 different output values can be achieved. Looking at the differences of its voltage levels in Fig. 2(a) it can be noticed that there is only one pair of closest values that are 222 and  122.


\begin{table}[h]
\caption{Resistance levels}
\label{table_example}
\begin{center}
\begin{tabular}{|c||c||c||c|}
\hline
  & M1$(\Omega)$ & M2 $(\Omega)$ & M3$(\Omega)$\\
\hline
0 & 232,068 & 232,132& 233,760\\
\hline
1 & 918,906& 749,831& 463,077\\
\hline
2 & 1537996,116& 498863,843& 26853,357\\
\hline
\end{tabular}
\end{center}
\end{table}

 Table 1 shows the different logical states within each of the three sub-cells (Fig. 1) to achieve 27 different levels.  Within a given sub-cell memristors are set to work in a ternary logic, with each logic state taking a different resistance value to ensure sufficient separation between the output voltages of the three sub-cells. It can be seen that resistance values for logic state 2 is extremely high which reflects the modeling issue of physical behaviour of memristors. This may be calibrated working with a real device, configuring the values of resistive network around the sub-cell.  

\begin{figure}[!ht]
\centering
\includegraphics[width=6cm]{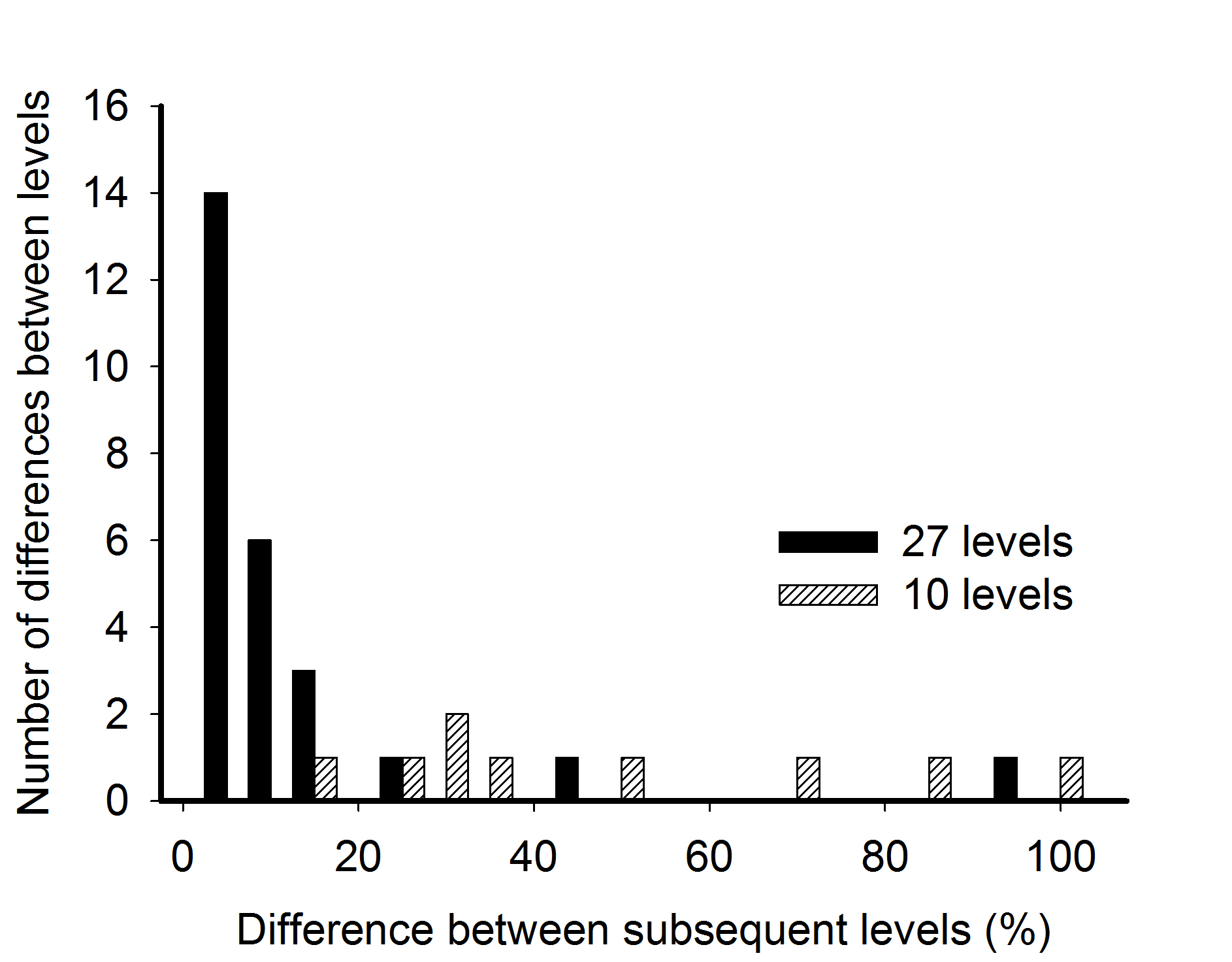}
\caption{The histogram of relative percentage difference between the subsequent levels.}
\end{figure}
{The histogram of differences between subsequent output levels like 222 and 122, 122 and 112 shown in Fig. 3(a) are given as a relative percentage in Fig. 4. It can be seen that an increase in the number of levels decreases the output voltage differences between the levels. For example in 27 level memory 23 subsequent levels have less than 20\% relative change in their values, while 10 level memory exhibits greater change with most of the values being between 20 and 40\%.This relative change indicates that a 10 level cell can be more robust in operation than a 27 level cell for the given set of programming voltage levels (i.e. [0V, 3V]).
} 

Table 2 on the other hand provides analysis on the sensitivity of the output states relative to the changes in write signal pattern applied through the write signal ($V_w$). The table indicates the results for a 5\% change in write signal $V_w$ pattern for 27 level and 10 level memory cell. { In the column corresponding to 10 level memory cell same values of different write patterns like 001, 010, 100 and 002, 020, 200 are highlighted with the same colors.} Overall, the 27 level cell exhibits higher degree of tolerance to changes in write signal pattern as opposed to 10 level cell, with average relative errors being 8.55\%, and 14.98\%, respectively.


\begin{table}[!ht]
\centering
\caption{Relative error for 5\% change in write signal pattern}
\label{my-label}
\begin{tabular}{p{2cm}|l|l}\hline
Write Pattern& \multicolumn{2}{c}{Relative error (\%)}\\\cline{2-3}
&27 levels&10 levels\\\hline
  222  & 89.22 $\pm$ 44.68 & 32.61 $\pm$ 13.38 \\
212 & 21.44 $\pm$ 3.25  &\cellcolor[HTML]{45F5CF} 53.72 $\pm$ 1.34  \\
122 & 17.20 $\pm$ 3.46  &\cellcolor[HTML]{45F5CF} 53.72 $\pm$ 1.34  \\
221 & 11.41 $\pm$ 0.33  &\cellcolor[HTML]{45F5CF} 53.72 $\pm$ 1.34  \\
211 & 12.77 $\pm$ 0.51  &\cellcolor[HTML]{C0B2F9} 30.60 $\pm$ 11.85 \\
112 & 11.52 $\pm$ 4.97  &\cellcolor[HTML]{C0B2F9} 30.60 $\pm$ 11.85 \\
121 & 9.16  $\pm$ 2.44  &\cellcolor[HTML]{C0B2F9} 30.60 $\pm$ 11.85 \\
220 & 0.18  $\pm$ 0.11  &\cellcolor[HTML]{F9F8B2} 5.70  $\pm$ 0.15  \\
210 & 4.36  $\pm$ 0.29  &\cellcolor[HTML]{F9B2B2} 5.57  $\pm$ 1.42  \\
120 & 13.25 $\pm$ 15.56 &\cellcolor[HTML]{F9B2B2} 5.57  $\pm$ 1.42  \\
202 & 1.66  $\pm$ 0.56  &\cellcolor[HTML]{F9F8B2} 5.70  $\pm$ 0.15  \\
111 & 7.72  $\pm$ 3.67  & 16.78 $\pm$ 10.97 \\
022 & 2.98  $\pm$ 0.67  &\cellcolor[HTML]{F9F8B2} 5.70  $\pm$ 0.15  \\
201 & 2.55  $\pm$ 0.03  &\cellcolor[HTML]{F9B2B2} 5.57  $\pm$ 1.42  \\
102 & 3.86  $\pm$ 1.15  &\cellcolor[HTML]{F9B2B2} 5.57  $\pm$ 1.42  \\
110 & 4.06  $\pm$ 2.82  &\cellcolor[HTML]{AFF5FC} 3.40  $\pm$ 2.28  \\
021 & 2.33  $\pm$ 0.29  &\cellcolor[HTML]{F9B2B2} 5.57  $\pm$ 1.42  \\
200 & 0.06  $\pm$ 0.03  &\cellcolor[HTML]{80FF80} 0.32  $\pm$ 0.07  \\
012 & 3.31  $\pm$ 0.90  &\cellcolor[HTML]{F9B2B2} 5.57  $\pm$ 1.42  \\
101 & 3.15  $\pm$ 1.16  &\cellcolor[HTML]{AFF5FC} 3.40  $\pm$ 2.28  \\
020 & 0.60  $\pm$ 0.26  &\cellcolor[HTML]{80FF80} 0.32  $\pm$ 0.07  \\
100 & 1.40  $\pm$ 0.97  &\cellcolor[HTML]{FFCCFF} 1.08  $\pm$ 0.74  \\
011 & 2.48  $\pm$ 0.85  &\cellcolor[HTML]{AFF5FC} 3.40  $\pm$ 2.28  \\
002 & 1.38  $\pm$ 0.14  &\cellcolor[HTML]{80FF80} 0.32  $\pm$ 0.07  \\
010 & 1.70  $\pm$ 0.03  &\cellcolor[HTML]{FFCCFF} 1.08  $\pm$ 0.74  \\
001 & 0.98  $\pm$ 0.01  &\cellcolor[HTML]{FFCCFF} 1.08  $\pm$ 0.74  \\
000 & 0.00  $\pm$ 0.00  & 0.00 $\pm$ 0.00  \\\hline
Average error  & 8.55  $\pm$ 3.30  & 14.98 $\pm$ 5.49 \\\hline
\end{tabular}
\end{table}


\section{Conclusions}
{Designing analog memory amongst vast possible applications of memristors draws extraordinary interest in neuromorphic engineering field. Taking into account the nanoscale size and fast switching property of memristors, memristive analog memory cells provide desirable means for building artificial neural networks. In this paper discrete analog memory is proposed using multiple memristive sub-cells. It is of note that increasing the number of memristive sub-cells of the memory results in increased number of analog output levels of the cell, therefore it can be concluded that continuous output levels can be achieved by combination of additional sub-cells.}

\bibliographystyle{IEEEtran}
\bibliography{scholar1}
\balance

\end{document}